\documentclass[useAMS,usenatbib]{mnras}

\usepackage{epsfig}
\usepackage{aas_macros}
\usepackage{natbib}
\usepackage{times}
\usepackage{graphicx, bm, amssymb, xcolor}
\usepackage{verbatim}
\usepackage[all]{hypcap}
\usepackage{xspace}
\usepackage{multirow}
\usepackage{pgf,tikz}
\usepackage{mathrsfs}
\usetikzlibrary{arrows}
\usepackage{amsmath}
\usepackage{subfigure}
\usepackage{array}

\usepackage{environ}
\makeatletter
\newsavebox{\measure@tikzpicture}
\NewEnviron{scaletikzpicturetowidth}[1]{%
  \def\tikz@width{#1}%
  \begin{lrbox}{\measure@tikzpicture}%
  \BODY
  \end{lrbox}%
  \pgfmathparse{#1/\wd\measure@tikzpicture}%
  \BODY
}
\makeatother

\newcommand{\be}{\begin{equation}}
\newcommand{\ee}{\end{equation}}
\newcommand{\bea}{\begin{eqnarray}}
\newcommand{\eea}{\end{eqnarray}}
\newcommand{\dd}{\mathrm{d}}

\title{On the physical mechanisms governing the cloud lifecycle in the Central Molecular Zone of the Milky Way \vspace{-0.2cm}}
\author{S.~M.~R.~Jeffreson$^1$\thanks{s.jeffreson@uni-heidelberg.de}, J.~M.~D.~Kruijssen$^1$, M.~R.~Krumholz$^2$ and S.~N.~Longmore$^3$ \\
$^1$Astronomisches Rechen-Institut, Zentrum f\"{u}r Astronomie der Universit\"{a}t Heidelberg, M\"{o}nchhofstra\ss e 12-14, 69120 Heidelberg, Germany \\
$^2$Research School of Astronomy \& Astrophysics, Australian National University, Canberra, ACT 2611, Australia \\
$^3$Astrophysics Research Institute, Liverpool John Moores University, 146 Brownlow Hill, Liverpool L3 5RF, UK
\vspace{-0.2cm}}

\hypersetup{draft}
\begin{document}

\date{Accepted 2018 May 1. Received 2018 April 9; in original form 2018 January 23. \vspace{-0.2cm}}

\pagerange{\pageref{firstpage}--\pageref{lastpage}} \pubyear{2017}

\maketitle

\label{firstpage}

\begin{abstract}
We apply an analytic theory for environmentally-dependent molecular cloud lifetimes to the Central Molecular Zone of the Milky Way. Within this theory, the cloud lifetime in the Galactic centre is obtained by combining the time-scales for gravitational instability, galactic shear, epicyclic perturbations and cloud-cloud collisions. We find that at galactocentric radii $\sim 45$-$120$~pc, corresponding to the location of the `$100$-pc stream', cloud evolution is primarily dominated by gravitational collapse, with median cloud lifetimes between $1.4$ and $3.9$~Myr. At all other galactocentric radii, galactic shear dominates the cloud lifecycle, and we predict that molecular clouds are dispersed on time-scales between $3$ and $9$~Myr, without a significant degree of star formation. Along the outer edge of the $100$-pc stream, between radii of $100$ and $120$~pc, the time-scales for epicyclic perturbations and gravitational free-fall are similar. This similarity of time-scales lends support to the hypothesis that, depending on the orbital geometry and timing of the orbital phase, cloud collapse and star formation in the $100$-pc stream may be triggered by a tidal compression at pericentre. Based on the derived time-scales, this should happen in approximately 20~per~cent of all accretion events onto the $100$-pc stream.
\end{abstract}

\begin{keywords}
Galaxy: centre --- stars: formation --- ISM: clouds --- ISM: evolution --- ISM: kinematics and dynamics --- galaxies: ISM
\vspace{-0.2cm}
\end{keywords}

\vspace{-0.2cm}
\section{Introduction} \label{Sec::introduction}
\begin{table*}
\begin{center}
  \caption{The dynamical time-scales used in the cloud lifetime theory of~\protect\cite{Jeffreson+Kruijssen18} and their physical interpretations.}
  \begin{tabular}{ m{1.3cm} m{1cm} m{8cm} m{2.5cm} m{2.5cm} }
  \hline
   Time-scale & Symbol & Physical meaning & Analytic form & Physical variables \\
  \hline
   $\tau_\kappa$ & $\kappa$ &
    Time-scale for the effect of epicyclic perturbations on GMCs. & 
    $\frac{4\pi}{\Omega \sqrt{2(1+\beta)}} \frac{1}{\sqrt{3+\beta}}$ & $\Omega$, $\beta$ \\
    $\tau_{\rm ff,g}$ & $f$ &
    Time-scale for the gravitational collapse of the ISM on approximately sub-Toomre length scales, as in~\cite{Krumholz+12}. & 
    $\sqrt{\frac{3\pi^2}{32\phi_P(1+\beta)}} \frac{Q}{\Omega}$ & $Q$, $\Omega$, $\beta$, $\phi_P$ \\
    $\tau_{\rm cc}$ & $c$ &
    Time-scale for collisions between GMCs~\citep{Tan00}. & 
    $\frac{2\pi Q}{9.4 f_G \Omega(1+0.3\beta)(1-\beta)}$ & $Q$, $\Omega$, $\beta$ \\
    $\tau_\beta$ & $\beta$ &
    Time-scale on which galactic shear pulls a cloud apart in the azimuthal direction. This is the only time-scale that has a fundamentally dispersive effect on molecular clouds. As such, the rate of galactic shear $\tau_\beta^{-1}$ is {\it subtracted} from the other rates in Equation~\ref{Eqn::cloud_lifetime}. & 
    $\frac{2}{\Omega(1-\beta)}$ & $\Omega$, $\beta$ \\
  \hline
  \end{tabular}
\end{center}
\label{Tab::time-scales}
\vspace{-0.25cm}
\end{table*}

The Central Molecular Zone (CMZ) of the Milky Way contains the largest concentration of high-density molecular gas in the Galaxy~\citep{Ferriere+07}. Despite this large gas reservoir, coupled with high gas pressures and velocity dispersions~\citep[e.g.][]{Oka+01}, the observed star formation rate (SFR) in the CMZ is $10$-$100$ times lower than that predicted by standard star formation relations~\citep{Yusef-Zadeh09,Immer+12,Longmore+13,Kauffmann+13,Barnes+17}. Galactic dynamical processes appear to play a dominant role in driving the evolution of the high-density clouds. This is supported by a growing body of observational evidence that star formation in the `$100$-pc stream' of gas at galactocentric radii of $\sim 100$~pc may be triggered by a tidal compression event, either at the pericentre of an eccentric orbit \citep{Longmore+13b,Rathborne+14a,Kruijssen+15,Henshaw+16b} or due to the change of the gravitational potential during accretion onto the inner CMZ \citep{Kruijssen+18b}. The global gas properties of the CMZ can be successfully reproduced by large-scale gas flows driven towards the central supermassive black hole (SMBH) by a combination of gravitational and acoustic instabilities, driving an episodic cycle of large-scale star formation and quiescence \citep{Kruijssen+14b,Krumholz+Kruijssen15,Krumholz+17}. The CMZ therefore presents a nearby example of the interplay between galactic dynamics, large-scale gas flows, the feeding of a central SMBH, star formation, and feedback. Its gas reservoir has similar properties to those observed in high-redshift galaxies~\citep{Kruijssen+Longmore13}, such that an understanding of the baryon cycle in our Galactic centre may also shed light on extragalactic star formation.

Throughout the Galaxy, giant molecular clouds (GMCs) host the majority of star formation~\citep{Kennicutt+Evans12}. In order to understand the baryon cycle in the CMZ, it is therefore necessary to understand its cloud-scale physics. In~\cite{Jeffreson+Kruijssen18}, we developed a theory for the cloud lifetime, dependent upon the large-scale dynamics of the galactic environment. Applied to the CMZ, our theory can be used to quantitatively predict the cloud lifetime and to understand the role played by galactic dynamics in cloud evolution and subsequent star formation. In this paper, we combine the analytic theory of~\cite{Jeffreson+Kruijssen18} with the model of~\cite{Krumholz+17}. We determine which large-scale dynamical processes are most important in setting the course of cloud evolution, and consequently star formation, in the gas inflow from radii of $\sim 500$~pc down to the $100$-pc stream. This not only gives a quantitative prediction for the variation in cloud lifetime with radius, but also divides the CMZ into dynamical regimes, in which cloud evolution is dominated by different dynamical processes. The dynamically-driven gas flows described in~\cite{Krumholz+17} must pass through each of these dynamical regimes on their way towards the central SMBH.

\section{Theory} \label{Sec::theory}
In~\cite{Jeffreson+Kruijssen18}, we introduced a theory for the molecular cloud lifetime, dependent upon the large-scale dynamics of the interstellar medium (ISM). Here we develop the salient points of this theory in relation to the central $500$~pc of the Milky Way. For a more detailed overview of the theory, we refer the reader to~\cite{Jeffreson+Kruijssen18}.

Our theory of the molecular cloud lifetime is independent of the size, structure and gravitational boundedness of molecular clouds, in accordance with the diverse range of objects that can observationally be classified as GMCs. It calculates the cloud lifetime as an environmentally-dependent quantity, consistent with the observed environmental dependence of the star formation efficiency per unit time in spiral and dwarf galaxies~\citep{Leroy+08}. Using only the observable properties of the ISM, the cloud lifetime $\tau$ in the CMZ can be quantified by adding the rates of the relevant large-scale dynamical processes in parallel, such that
\begin{equation}
\label{Eqn::cloud_lifetime}
\tau = |(\tau_\kappa^{-1}+\tau_{\rm ff,g}^{-1}+\tau_{\rm cc}^{-1}-\tau_\beta^{-1})|^{-1}.
\end{equation}
The different time-scales in this equation and their physical variables are summarised in Table~\ref{Tab::time-scales}, where $\Omega$ is the angular velocity of the midplane ISM, $\beta$ is the galactic shear parameter
\begin{equation}
\label{Eqn::beta}
\beta = \frac{\dd \ln{v_c}}{\dd \ln{R}},
\end{equation}
for circular velocity $v_c(R)$ at galactocentric radius $R$, and $Q$ is the \citet{Toomre64} $Q$ parameter for the ISM midplane gas
\begin{equation}
\label{Eqn::Toomre_Q}
Q = \frac{\kappa \sigma_g}{\pi G \Sigma_g},
\end{equation}
for an epicyclic frequency $\kappa$, midplane gas velocity dispersion $\sigma_g$, and midplane gas surface density $\Sigma_g$. The variable $\phi_P$ quantifies the contribution of the stellar potential to the ISM pressure, as defined in~\cite{Elmegreen89},
\begin{equation}
\label{Eqn::phi_P}
\phi_P = 1+\frac{\Sigma_s}{\Sigma_g} \frac{\sigma_g}{\sigma_s},
\end{equation}
where $\sigma_s$ and $\Sigma_s$ refer to the stellar velocity dispersion and surface density, respectively. The variable $f_G=0.5$ in $\tau_{\rm cc}$ is a `collision probability' parameter defined and fitted to observations in~\cite{Tan00}. All time-scales depend inversely on the angular velocity $\Omega$, such that the normalised cloud lifetime $\tau/\Omega^{-1}$ is described within a parameter space spanned by the four physical variables $\beta$, $Q$, $\phi_P$ and $f_G$. Of these, we fix $f_G$ to its above fiducial value, and note that only the time-scale $\tau_{\rm ff,g}$ for gravitational free-fall depends weakly on $\phi_P$. The cloud lifetime therefore varies within a fundamental parameter space spanned by $\beta$, $Q$ and $\Omega$, with a secondary dependence on $\phi_P$. Values of these variables for the CMZ are accessible through measurements of its rotation curve, velocity dispersion profile, and surface density profile. Since neither the Galactic bar nor the Galactic spiral arms extend down to the maximum galactocentric radius of $\sim 500$~pc considered here, we have excluded the dynamical time-scale $\tau_{\Omega_{\rm P}}$ for spiral-arm crossings, although it is discussed in~\cite{Jeffreson+Kruijssen18}.

\vspace{-0.4cm}
\section{Application to the CMZ} \label{Sec::application_CMZ}
In order to use Equation~\ref{Eqn::cloud_lifetime} to calculate the cloud lifetime in the CMZ, we require its rotation curve, velocity dispersion profile and surface density profile. The accurate measurement of velocity dispersions in the CMZ is currently an active topic of research~\citep{Shetty+12,Henshaw+16a,Henshaw+16b}, while the edge-on CMZ viewing angle prohibits the acquisition of an accurate face-on surface density profile. As such, we use observational data for the rotation curve of the CMZ from~\cite{Launhardt+02}, but use the gas velocity dispersions and gas surface densities produced by simulation run m10r050f10 from the dynamical model of~\cite{Krumholz+17}, which are consistent with the gas properties inferred observationally for the CMZ \citep[see e.g.~their Figures~10 and~14 and compare to][]{Kruijssen+14b,Henshaw+16a,Henshaw+16b}. This numerical simulation successfully reproduces several of the observed properties of the CMZ, in particular the large-scale gas distribution.

\begin{figure}
  \label{Fig::timescales_485}
    \includegraphics[width=\linewidth]{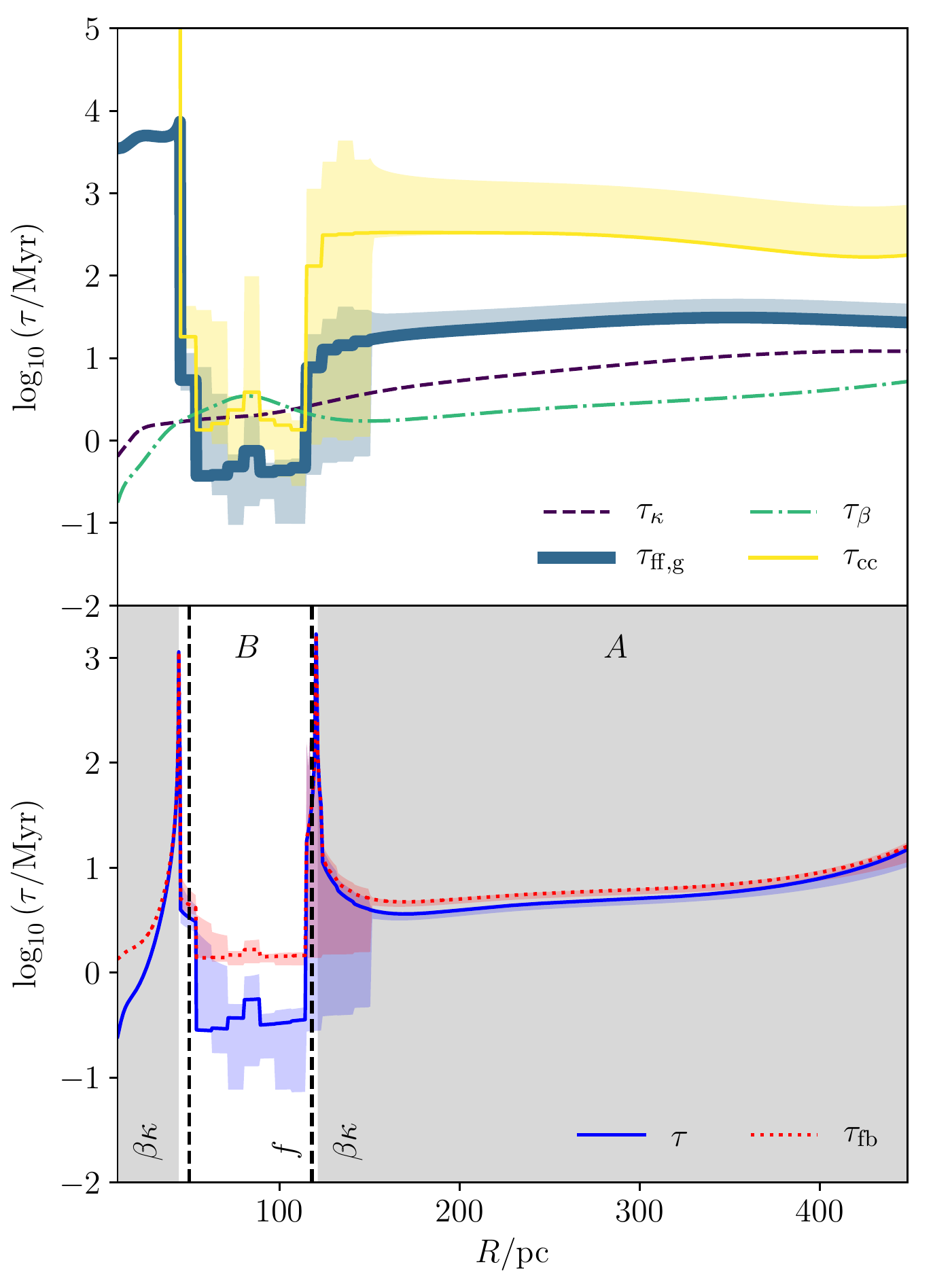}
    \vspace{-0.5cm} \caption{Values of the cloud lifetime predicted by our model for the current gas distribution in the best-matching simulation of~\protect\cite{Krumholz+17}, including the standard deviation in the model-dependent time-scales across the last $100$~Myr of all models (see the text). The upper panel gives each of the time-scales of dynamical evolution. The blue solid line in the lower panel gives the lifetime calculated using Equation~\ref{Eqn::cloud_lifetime}, and the red dotted line adds a feedback time-scale $\tau_{\rm fb} \sim 1.1$~Myr to this lifetime (see the text). The grey shaded areas indicate the galactocentric radii at which the rate of galactic shear $\tau_\beta^{-1}$ is higher than the combined rates of all other mechanisms, while the black dashed vertical lines delineate the {\it regions of relevance}, discussed in Section~\ref{Sec::application_CMZ} and labelled according to Table~\ref{Tab::time-scales}. The labels `$A$' and `$B$' refer to the outer CMZ and the vicinity of the $100$-pc stream, respectively. \vspace{-0.2cm}}
\end{figure}

In Figure~\ref{Fig::timescales_485}, we display the time-scales of each cloud evolutionary mechanism (top panel) and the resulting cloud lifetimes (bottom panel) as a function of galactocentric radius, at a simulation time of 485 Myr in the model m10r050f10, corresponding to the gas properties that best match those observed at the current epoch. We also display the standard deviation in each quantity at each radius over the whole range of model parameters during the most recent $100$~Myr in~\cite{Krumholz+17}, to provide an indication of how much they vary.\footnote{The standard deviation is computed by sampling the last $100$~Myr of evolution at $5$~Myr intervals for the models m01r050f10, m03r025f10, m03r050f05, m03r050f10, m03r050f20 and m10r050f10.} The value of $\phi_P$ has been calculated at each galactocentric radius using the stellar velocity dispersion of $\sigma_{\rm s} \approx 100$~kms$^{-1}$ from~\cite{deZeeuw93}, the rotation curve from~\cite{Launhardt+02}, and the gas surface density and velocity dispersion profiles from~\cite{Krumholz+17}. As in~\cite{Jeffreson+Kruijssen18}, we indicate {\it regions of relevance}, enclosed by black dashed lines. The {\it relevance} of a single cloud evolutionary mechanism depends on the ratio of its time-scale to the minimum evolutionary time-scale $\tau_{\rm min}$ (where $\tau < \tau_{\rm min}$), or to the cloud lifetime $\tau$ (where $\tau > \tau_{\rm min}$ due to shear support). If this ratio exceeds a value of $2$, i.e.~the mechanism occurs at under half the rate of the dominant evolutionary mechanism for $\tau < \tau_{\rm min}$, or at under half the rate of cloud destruction for $\tau > \tau_{\rm min}$, then its effect on cloud evolution is deemed {\it irrelevant}.

We find that the CMZ can be divided into two distinct regimes, corresponding to the grey- and white-shaded areas in the bottom panel of Figure~\ref{Fig::timescales_485}. The grey-shaded areas ($R \ga 120$~pc and $R \la 45$~pc) indicate the galactocentric radii that are dominated by galactic shear, to the extent that the rate of shearing outpaces the combined rates of all dynamically-compressive cloud evolutionary mechanisms ($\tau_\beta^{-1} > \tau_\kappa^{-1} + \tau_{\rm ff,g}^{-1} + \tau_{\rm cc}^{-1}$). The white-shaded area ($45 \la R/{\rm pc}\la 120$) indicates the radii that are dominated by dynamically-compressive mechanisms of cloud evolution. Due to the extremely low gas column density at galactocentric radii $R \la 45$~pc, we will ignore the innermost grey-shaded area from hereon. We will identify the shear-dominated regime with the outer CMZ (labelled `$A$') and will identify the dynamically-compressive regime with the radii close to the $100$-pc stream (labelled `$B$').

In the vicinity of the $100$-pc stream (`$B$', $45 \la R \la 120$~pc in Figure~\ref{Fig::timescales_485}), the majority of GMCs are expected to collapse and form stars, due to the dominance of dynamically-compressive cloud evolutionary mechanisms. At radii from $50$ to $110$~pc, gravitational free-fall `$f$' is the only relevant mechanism of cloud evolution, leading to short cloud lifetimes between $0.3$ and $5$~Myr at the current epoch. The only exception to the dominance of gravity arises at the entrance to the $100$-pc stream at $\sim 120$~pc, where the volume density of gas entering the star-forming ring is still low enough that epicyclic perturbations and galactic shear compete with gravitational free-fall (note the equality of all time-scales at $\sim 120$~pc in the top panel of Figure~\ref{Fig::timescales_485}). The importance of epicyclic perturbations at $\sim 120$~pc is consistent with the hypothesis that star formation in the $\sim 100$~pc stream may be triggered by tidal compressions due to pericentre passages of molecular clouds on epicyclic orbits~\citep{Longmore+13b,Kruijssen+15,Henshaw+16b}. In the outer CMZ `$A$', cloud lifetimes are consistently longer, around $10$~Myr, due to the increased degree of shear support, which balances closely with the dynamically-compressive mechanisms of cloud evolution. At these radii, self-gravity is irrelevant and thus cloud evolution is controlled by shear and epicyclic perturbations, `$\beta \kappa$'. We expect the majority of GMCs in the outer CMZ to have low star formation efficiencies per unit mass, and eventually to be dispersed by galactic shear.

As indicated by the standard deviations in $\tau_{\rm ff, g}$, $\tau_{\rm cc}$ and the cloud lifetime $\tau$, the results do not vary much over the past $100$~Myr (despite considerable variations in the SFR), nor do they depend strongly on the parameter choices of the~\cite{Krumholz+17} model. That is, gravitational free-fall always dominates in the vicinity of the $100$-pc stream, and a combination of galactic shear and epicyclic perturbations dominate elsewhere. The cloud lifetimes themselves vary by 3~per~cent at $\sim 200$~pc (region A) and by up to 30~per~cent at $\sim 100$~pc (region B), which corresponds to a shift from $0.3$ to $0.2$~Myr in the minimum cloud lifetime for region B. The major downward uncertainty between $120$--$150$~pc arises because region B extends to $\sim 150$~pc in a small subset of the complete range of models considered.

In the bottom panel of Figure~\ref{Fig::timescales_485}, we also include the feedback-adjusted cloud lifetime. In the model of \citet{Jeffreson+Kruijssen18}, we assume that the lifetime of a cloud is determined by its evolution towards star formation, and that destruction by feedback occurs on a much shorter time-scale. This assumption is appropriate for galactic discs, where the dynamical time-scales generally greatly exceed the time-scale for gas removal by feedback, but breaks down in the $100$-pc stream, where the dynamical time-scales are short. We thus add a feedback time-scale of $\tau_{\rm fb}=1.1~{\rm Myr}$ to the calculated cloud lifetime, which corresponds to the time taken to traverse the section of the 100-pc stream between Sgr B2, where stellar feedback first sets in, and Sgr B1, where the most of the molecular gas has been blown out~\citep{Kruijssen+15,Barnes+17}. With the addition of this feedback time-scale, the range of cloud lifetimes predicted for the $100$-pc stream is raised to $1.4$--$3.9$~Myr. We emphasise that the feedback time-scale is the only result in this paper that depends on an evolutionary progression of cloud evolution along the $100$-pc stream, from Sgr~B2 to Sgr~B1.

\begin{figure}
  \label{Fig::lifetime_map_alltimes}
  \includegraphics[width=.99\linewidth]{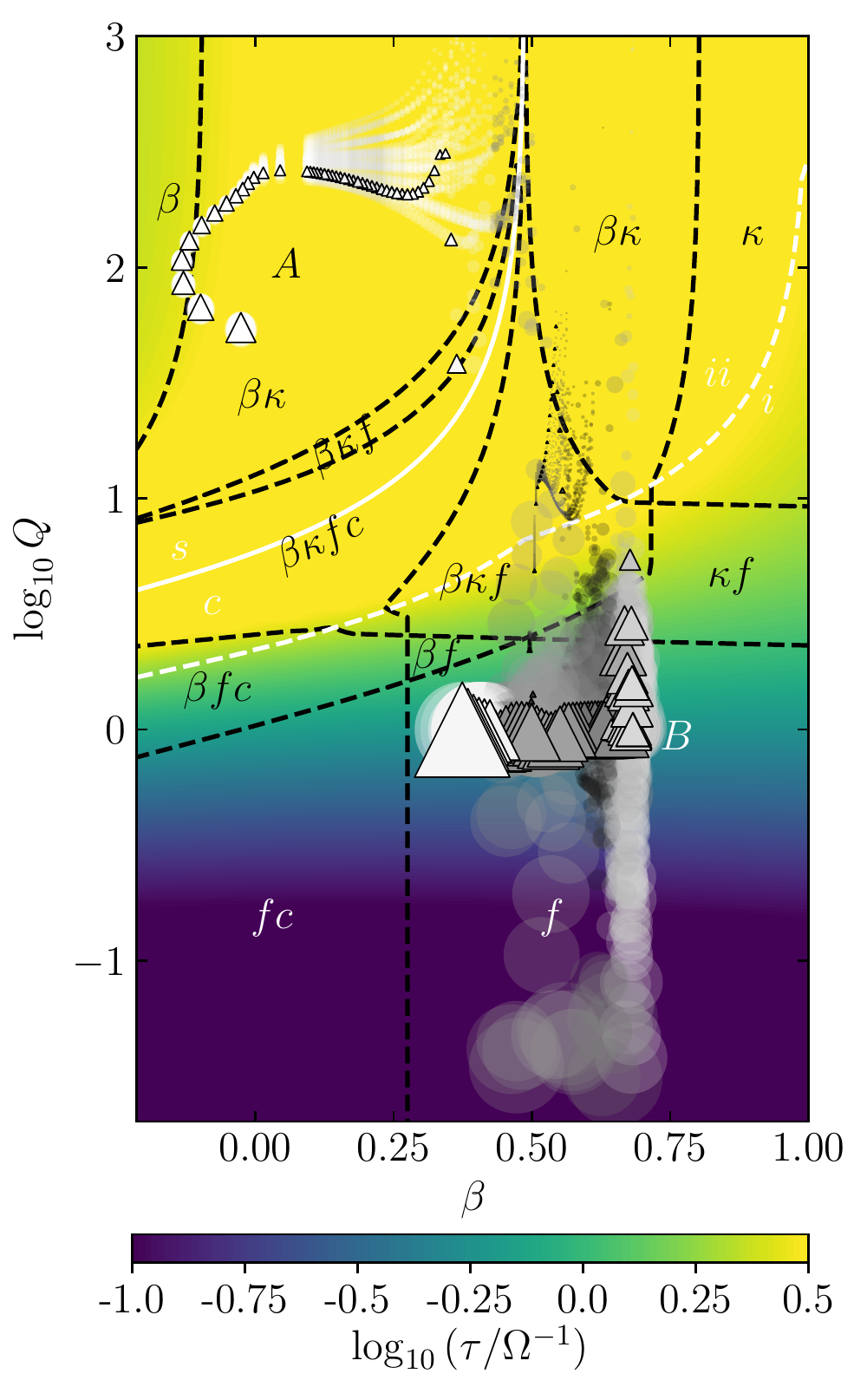}
  \vspace{-0.4cm} \caption{Values of the predicted cloud lifetime in units of angular velocity $\Omega$ (coloured contours) for a cross-section of $(\beta,\log{Q})$ parameter space with $\phi_P\sim 1$, overlaid with values of $\beta$ and $\log{Q}$ (translucent circles) for the CMZ, at $10$~Myr intervals from $100$ to $500$~Myr in the best-matching simulation of~\protect\cite{Krumholz+17}. The data at $485$~Myr (the current epoch) are given by triangles. The data points are colour-coded on a grey scale between black and white by galactocentric radius, where a lighter colour corresponds to a larger radius. Their area is weighted on a linear scale by the total gas mass in each radial interval, where the largest data points represent the largest quantities of molecular gas. The dashed black lines enclose the \textit{regions of relevance} for each time-scale, labelled as in Figure~\ref{Fig::timescales_485}. The dashed white lines divide the regimes `$ii$' in which the cloud lifetime is longer than the minimum evolutionary time-scale from the regimes `$i$' in which it is shorter. The solid white lines divide the shear-dominated regime `$s$' from the dynamically-compressive regime `$c$' (grey- and white-shaded regions, respectively, in Figure~\ref{Fig::timescales_485}). The labels `$A$' and `$B$' denote the regions of the CMZ shown in Figure~\ref{Fig::timescales_485} and discussed in Section~\ref{Sec::application_CMZ}. \vspace{-0.2cm}}
\end{figure}

In Figure~\ref{Fig::lifetime_map_alltimes}, we show the CMZ model by \citet{Krumholz+17} in the fundamental parameter space $(\beta, Q)$, over which the normalised cloud lifetime, $\tau/\Omega^{-1}$, varies. Triangular points denote the data at the current epoch, corresponding to the cloud lifetimes in Figure~\ref{Fig::timescales_485}. Circular points denote the data at all other epochs. The values of $\beta$ are taken from the observed rotation curve, which is assumed to be constant in time, and the {\it regions of relevance} are delineated by black dashed lines. The shear-dominated regime `$s$', corresponding to the grey-shaded regions in Figure~\ref{Fig::timescales_485}, is separated by a white solid line from the dynamically-compressive regime `$c$'. We set $\phi_P \sim 1$, the value appropriate to the $100$-pc stream~\citep[c.f.][]{deZeeuw93,Launhardt+02,Krumholz+17}, because it only affects $\tau_{\rm ff,g}$ and the effect of gravitational free-fall is strongest in the $100$-pc stream. We note that higher values of $12 \la \phi_P \la 100$, appropriate to radii of $120$--$500$~pc, do not significantly alter the predicted cloud lifetime outside the $100$-pc stream.

We find that the two distinct regimes `$A$' and `$B$', corresponding to different galactocentric radii in Figure~\ref{Fig::timescales_485}, are also distinct in $(\beta, Q, \Omega)$ parameter space. The outer (`$A$', $\ga 120$~pc) region of the CMZ, through which gas flows inwards with large scale heights and low volume densities, is characterised by an approximately-flat rotation curve and very high levels of gravitational stability ($\beta < 0.5$ and $Q \ga 60$). According to the model of~\cite{Krumholz+17}, `$A$' can be interpreted as a body of gas spiralling in towards the star-forming, ring-shaped stream `$B$' at $\sim 100$~pc, propelled by shear-driven acoustic instabilities. Due to high levels of shearing and gravitational stability, GMCs in the outer CMZ fall exclusively in the dynamically-dispersive regime `$s$' of parameter space, and are governed by a combination of galactic shear and epicyclic perturbations `$\beta \kappa$'. The competition between dynamically-compressive and dynamically-dispersive mechanisms of cloud evolution in this regime elongates the cloud lifetime to between $2.3$ and $3.3$ orbital times $1/\Omega$. Note that we have excluded $>3 \sigma$ outliers for this and all subsequent ranges, in order to reflect the typical cloud lifetime at each interval of galactocentric radii.

In the vicinity of the $100$-pc stream `$B$', between galactocentric radii of $45$ and $120$~pc, the gas inflow stalls due to a local shear minimum in the rotation curve~\citep{Krumholz+Kruijssen15}, where $0.5 \la \beta \la 0.75$. It condenses to small scale heights ($\sim 3$~pc) and high volume densities ($\sim 500$~$M_\odot {\rm pc}^{-3}$), such that the level of gravitational stability falls as low as $Q \sim 0.1$. At these radii, we expect that the majority of GMCs are governed by gravity alone (in regime `$f$' of parameter space), and are therefore destroyed by gravitational collapse and the subsequent stellar feedback. In particular, we expect cloud lifetimes on the outside of the $100$-pc stream (between $100$ and $120$~pc, corresponding to the lighter-coloured data points in region `$B$') to be very short ($\sim 1$ orbital time $1/\Omega$). On the inside of the stream (between $45$ and $100$~pc, corresponding to the darker-coloured points in region `$B$'), the molecular gas surface density is depleted by star formation~\citep{Krumholz+17}, leading to gas masses as low as one $1000$th of the mass on the outside of the stream (compare the areas of the dark- and light-coloured data points in region `$B$' of Figure~\ref{Fig::lifetime_map_alltimes}). Due to its low gas fraction, this remaining material has a high degree of gravitational stability, leading to longer cloud lifetimes (between $1$ and $3$ orbital times).

The molecular gas that survives the star-forming ring exits the local shear minimum and makes its way towards the nuclear cluster and eventually the central SMBH~\citep{Krumholz+17}. This inner region of the CMZ is not shown in Figure~\ref{Fig::lifetime_map_alltimes}, because the vast majority of molecular gas in the~\cite{Krumholz+17} model is either consumed by star formation or blown out by feedback at earlier times in the star-forming ring, producing unreliable values of $Q$ at $\la 45$~pc. However, we do expect that the cloud lifecycle in the inner CMZ ($R \la 45$~pc) is controlled by similar mechanisms as in the outer CMZ `$A$' due to its flat ($\beta<0.5$) rotation curve, but with higher levels of gravitational stability due to its even lower gas density.

Comparing the triangular (current, quiescent phase) and circular (all snapshots) data points in Figure~\ref{Fig::lifetime_map_alltimes}, it is clear that there is little time-variation of the parameters $\beta$ and $Q$ between the starburst and quiescent phases of the~\citet{Krumholz+17} model, resulting in CMZ cloud lifetimes that are also relatively time-invariant. Over a period of $\sim 500$~Myr in simulation time, the radial extent of the gravity-dominated regime `$c$' and the radial extents of each region of relevance are constant to within 10~per~cent. This is a direct result of the shape of the gravitational potential and, hence, the rotation curve. As the CMZ evolves, the time-scales on which gravity-dominated clouds are destroyed by collapse and feedback in the $100$-pc stream `$B$', and on which shear-dominated clouds are dispersed by galactic shear outside the $100$-pc stream `$A$', is relatively constant. The most notable exception to time-invariance occurs in the vicinity of the $100$-pc stream `$B$', where a sharp drop in the cloud lifetime occurs during each starburst phase, corresponding to the scatter of points down to $Q \sim 0.1$ in regime `$f$' of Figure~\ref{Fig::lifetime_map_alltimes}. This is due to the simultaneous drop in the turbulent velocity dispersion and the rise in the molecular gas surface density that accompanies a starburst, producing a sudden drop in Toomre $Q$.

Although gravitational free-fall is the only relevant cloud evolutionary mechanism throughout the majority of the $100$-pc stream (see regime `$f$' of Figure~\ref{Fig::lifetime_map_alltimes}) a number of data points are also located in regimes `$\kappa f$' and `$\beta \kappa f$', where we expect epicyclic perturbations to have a significant influence on cloud evolution. This is consistent with the hypothesis that tidal compressions may trigger cloud collapse and star formation in the $100$-pc stream, due to pericentre passages of molecular clouds on eccentric orbits, or by accretion onto the stream~\citep{Longmore+13b,Kruijssen+18b}.

We may estimate the fraction of clouds whose collapse is triggered by epicycles, by first calculating the fraction of clouds $N(R)$ that survive until they reach galactocentric radius $R$ in the gravity-dominated regime `$c$' (i.e.~in the vicinity of the $100$-pc stream). This is given by the differential equation
\begin{equation}
\label{Eqn::rate_eqn}
v_R \frac{\dd N(R)}{\dd R} = -\tau^{-1}(R) N(R),
\end{equation}
where $v_R$ is the radial inflow velocity, calculated self-consistently at each radius in the model of~\cite{Krumholz+17}, and $\tau^{-1}(R)$ is the rate of cloud destruction, as calculated in Equation (\ref{Eqn::cloud_lifetime}). Solving this equation numerically for $N(R)$, we find that at the current epoch, 100~per~cent of clouds are destroyed between $R=120$~pc and $R=115$~pc, at the very outer edge of the gravity-dominated regime. That is, the inflow velocity of the clouds drops significantly as they spiral inwards towards the local shear minimum, so that all are destroyed before their apocentric radii shrink from $120$ to $115$~pc. This demonstrates that the groups of darker-coloured points (indicating smaller radii) in regimes `$\kappa f$' and `$\beta \kappa f$' of Figure~\ref{Fig::lifetime_map_alltimes} do not give meaningful information about the course of cloud evolution: for radii close to $45$~pc, very few molecular clouds remain.

Of the clouds that are destroyed in each interval of galactocentric radius, we may estimate the statistical fraction $F(R)$ of cloud destruction through star formation and feedback that is driven by epicycles, i.e.~the fraction of star formation events that is driven by pericentre passages. This is defined to exclude cloud dispersal by shear and (in regime `$c$' of parameter space) is given by
\begin{equation}
F(R) = \frac{\tau_\kappa^{-1}}{\tau_\kappa^{-1} + \tau_{\rm ff,g}^{-1} + \tau_{\rm cc}^{-1}},
\end{equation}
The overall fraction of clouds $F$ that are destroyed by epicycles is then given by the product of $F(R)$ and the fraction of clouds destroyed between $R$ and $R+\Delta R$, summed over all radii, so that
\begin{equation}
F = \sum_{R}{[N(R+\Delta R)-N(R)] F(R)} \approx 0.2,
\end{equation}
at the current epoch. That is, 20~per~cent of cloud destruction, upon accretion onto the $100$-pc stream, is caused by epicyclic perturbations (i.e.~pericentre passages). Across the full duty cycle at times close to the current epoch, the fraction of cloud destruction caused by epicyclic perturbations does not vary significantly and remains (with mostly downward variations) in the range 10--30~per~cent.

\vspace{-0.25cm}
\section{Conclusions} \label{Sec::conclusions}
Using the theoretical output of~\cite{Krumholz+17}, along with the observed rotation curve in the CMZ of the Milky Way, we have applied the theory of cloud lifetimes presented in~\cite{Jeffreson+Kruijssen18} to the Galactic centre.

From a cloud evolutionary perspective, we find that the CMZ is divisible into two dynamical regimes. At galactocentric radii from $\sim 120$-$500$~pc, the cloud lifecycle is primarily dominated by galactic shear, to the extent that the rate of shearing is faster than the combined rates of all other cloud evolutionary mechanisms. At these galactocentric radii we expect clouds to be sheared apart on time-scales between $3$ and $9$~Myr, before collapse and star formation can occur, leading to low star formation efficiencies. Conversely, at galactocentric radii from $\sim 45$-$120$~pc, we expect to find clouds that collapse and form stars on much shorter time-scales, with median lifetimes between $\sim 0.3$ and $2.8$~Myr. If we lift the assumption of instantaneous stellar feedback and include a gas removal time-scale of $\tau_{\rm fb}=1.1~{\rm Myr}$, motivated by observations, this range of cloud lifetimes becomes 1.4--3.9~{\rm Myr}.

At the outer edge of the $100$-pc stream, the time-scale for epicyclic perturbations, which quantifies the influence of orbital eccentricity on the cloud lifecycle, obtains equality with the free-fall time-scale. This result is consistent with the hypothesis of tidally-triggered collapse in the $100$-pc stream, as initially proposed by~\cite{Longmore+13b} and later expanded by~\citet{Kruijssen+15} and~\citet{Henshaw+16b}. While the similarity of the free-fall and epicyclic time-scales implies that some accreting gas streams may collapse due to a tidal compression at pericentre (in approximately 20~per~cent of cases), it also means that some gas streams may undergo free-fall collapse due to their arrival on the $100$-pc stream, before pericentre is reached (in around 80~per~cent of cases). This simple time-scale argument corroborates the numerical simulations presented by~\citet{Kruijssen+18b}, who show that the compressive tidal field at the radii of the $100$-pc stream may have an equal, if not stronger effect on cloud evolution than the pericentre passage. Since the gas is flowing in from larger radii, both collapse mechanisms lead to an evolutionary progression of star formation, either post-pericentre, or after their moment of accretion onto the $100$-pc stream. In combination with the known orbits of the CMZ gas streams, this provides an absolute evolutionary timeline that allows the cloud lifetimes predicted here to be directly tested with currently available observations.

\vspace{-0.25cm}
\section*{Acknowledgements}
SMRJ and JMDK acknowledge funding from the German Research Foundation (DFG) in the form of an Emmy Noether Research Group (grant number KR4801/1-1). JMDK acknowledges support from the European Research Council (ERC) under the European Union's Horizon 2020 research and innovation programme via the ERC Starting Grant MUSTANG (grant number 714907). MRK acknowledges support from the Australian Research Council’s Discovery Projects funding scheme (project DP160100695). We thank an anonymous referee for a constructive report, which improved the paper.

\vspace{-0.25cm}
\bibliographystyle{mnras}{}
\bibliography{bibliography}

\bsp

\label{lastpage}

\end{document}